\documentstyle[sprocl,epsfig]{article}
 
\input{psfig}
\def\Journal#1#2#3#4{{#1} {\bf #2}, #3 (#4)}
 
\def\NPB{{\em Nucl. Phys.} B}
\def\PLB{{\em Phys. Lett.}  B}
\def\PRL{\em Phys. Rev. Lett.}
\def\PRD{{\em Phys. Rev.} D}

\def\be{\begin{equation}}
\def\ee{\end{equation}}
\def\bea{\begin{eqnarray}}
\def\eea{\end{eqnarray}}
\begin{document}
 
\title{Wilson line in high temperature particle physics }
 
\author{ S. Bronoff}
 
\address{Centre Physique Th\'eorique au CNRS, Case 907, Luminy\\ F13288,
Marseille Cedex, France}
 
\author{G. Dvali}
 
\address{CERN Theory Division, CERN\\ 1211 Geneva 23,Switzerland}
 
\author{K. Farakos}
 
\address{National Technical University, Physics Dept., Zografou Campus\\ GR 15773, Athens, Greece}
 
\author{C.P. Korthals Altes}
 
\address{Centre Physique Th\'eorique au CNRS, Case 907, Luminy\\ F13288,
Marseille Cedex, France}

 
\maketitle\abstracts{
The physics of the Wilson line leads to new developments in high temperature particle physics. The main tool is the effective action for a given
fixed value of the phase of the Wilson line. It furnishes a 
gauge invariant infrared cut off, and yields for small values of the phases 
a systematic procedure for obtaining a power series in the coupling g and glog(1/g). It breaks the centergroup symmetry of the gauge group only at high temperature 
so  leads to domain walls disappearing at low temperatures. It shows long lived
 metastable states in the standard model, SU(5), SO(10) and its SUSY partners, with possibilities for CP violation
and thermal inflation.}
  
\section{Introduction}
 
Phase transitions have been a recurrent theme at this meeting.
We have heard about the quantitative progress in understanding the electro-weak
transition~\cite {laine} and along similar lines the SU(5) GUT
phase transition~\cite{rajantie} and to some extent the deconfinement transition~\cite{rummukainen}.  
 
Despite their differences there is a common feature: all of them have an
order parameter, the Wilson line. The Wilson line is strictly speaking an
order parameter, when there are no complex representations of the gauge
group present in the particle content. Only then it will be strictly zero in
one phase and non-zero in the other.

Usually the Wilson line is mentioned in discussions of the quark gluon
transition (apart from the chiral condensate), not of the other transitions, like the electro-weak and GUT
transitions where the Higgs fields are the protagonists in the transition.

In this paper we will discuss the role of the loop; in particular its 
phase, and the dependence of the free energy on that phase. The latter is
called the effective action and is the main mathematical tool.

For small values of the phases this effective action is a gauge invariant infra red cut off version of the free energy and the Debije mass. As such it is a nice starting point for the perturbative evaluation of these quantities~\cite{debbio}. Infra red singularities upset the usual series in $g^2$, and their computation necessitates a cutoff. This is briefly discussed in section~\ref{sec:infra}. 

For values of the phase in the centergroup one finds metastable minima.
They have one common property: they are very long lived on any cosmological scale. This is true for the standard model and for GUT models like
SU(5) and SO(10) and their SUSY versions. A rather detailed description of
where the minima are and how they decay is found in section~\ref{sec:sm}. 
Then we turn to the other order parameters, the Higgs fields. The potential has  qualitatively different behaviour due to the Wilson line condensate, in  case the Higgs field carries Z(N) charge. This is the subject of section~\ref{sec:higgs}.

In section~\ref{subsec:phys}
we turn to the physics of these metastable states. As the CP violating
properties of some of the metastable states have been analysed elsewhere
\cite{watson}, we limit ourselves to thermal inflation. This happens in the SO(10) model and its SUSY analogue. 

 The question of how the universe can arrive in such a 
metastable state arises naturally~\cite{watson}.  There has been debate on this~\cite{gavela} , as well as on the thermodynamical properties of the metastable states~\cite{semenoff}
and the physical relevance of the phase of the loop \cite{smilga}. On the 
latter two the reader can find  partial satisfaction in section~\ref{sec:order}. 
 
\section{Wilson line as order parameter}\label{sec:order}

In this section we will resume some well known facts about gauge
systems at high temperature in equilibrium.
First the academic but instructive case of pure SU(N) theory is reviewed.
Then we introduce the Wilson line as the order parameter that describes walls, and the effective action that controlls its behaviour.
In the last subsection we discuss the boundary conditions that trigger
localised walls.

\subsection{Pure gauge theory}\label{subsec:sun}

Let us consider 
 a pure SU(N) gauge theory~\cite{thooft}.
 
Take an elongated box in 3D with size $L_{tr}^2L_z$. The size in the z-direction
 is by far
the largest in order to have walls that separate vacuum states. Boundary conditions on the gauge potentials are periodic.

 Physical states are
by definition invariant under periodic gauge transformations.This means Gauss' law is satisfied everywhere. Consider now a gauge transformation periodic in the transverse directions, but periodic modulo
a centergroup element $\exp{i{2k\pi\over N}}$ in the z-direction. Such transformations $U_k$  differ from one another by a periodic transformation,
so have {\it{all}} the same effect on a physical state. Like a periodic transformation they do commute with the Hamiltonian. So we can diagonalise them simultaneously. Acting on the physical states they have eigenvalues
 all of which are again Z(N) phases.\footnote{That's because $U_k^N$
is periodic, hence does not change the physical state.}

 We can easily get a physical picture of what these
states are: start from a state on which {\it {all}} operators $U_k$ have eigenvalue one. This state has no electric flux, only glueballs are thermally
excited. Now take a string running from $z=0$ to 
$z=L_z$ given by
the  path-ordered operator $TrP\exp{i\int dz A_z(x,y,z)}$ acting on the no flux
state. This state has one electric flux, since the eigenvalue of $U_k$ is
$\exp{ik{2\pi\over N}}$. If we add another string  $k$ will be replaced by $2k
$ in the phase.  Note that a state with a given flux $e$ can have e mod N strings.

We can create strings independently in all directions. This means that a state will by given by a flux vector $\vec e$. Such a flux  state is obtained by acting with a projector
$P_{\vec e}$ on a physical state. This projector is related to the operators
$U_{\vec k}$ by a Z(N) Fourier transform:
\begin{equation}
P_{\vec e}={1\over N^3}\sum_{\vec k}\exp{-i\vec k.\vec e{2\pi\over N}}U_{\vec k} 
 \label{eq:proj}
\end{equation}

 Since flux is conserved (the Hamiltonian commutes with the $U_{\vec k}$), we can define
a flux free energy $F_{\vec e}$ by the Gibbs trace over flux $\vec e$ states:
\begin{equation}
\exp{-F_{\vec e}\over T}\equiv \sum\langle \vec e\vert\exp{-H\over T}\vert \vec e \rangle
\label{eq:free}
\end{equation}
These flux free energies should tell us how the strings behave. 
Now we expect at low T confinement, hence a string tension $\rho$. Let us take the case of one flux in the z-direction. Then at low T ($\ll \rho$):
\begin{equation}
F_1-F_0=\rho L_z
\label{eq:string}
\end{equation}

For high enough T($T_c$) this behaviour changes into:

\begin{equation}
F_1-F_0=L_z\exp{-\alpha(T)\over T L_{tr}^2}
\label{eq:surface} 
\end{equation}
The physics of these equations is simple. At low T
creation of one string has a small probability $\sim\exp{-{\rho\over T} L_z}$.
At the same time the difference between the free energies increases as the length of the box increases. The unique groundstate is given by $F_{\vec 0}$.

 As the temperature goes up
so goes the probability for exciting strings. The number of strings present in the Gibbs sum grows
and so does the entropy. At $T_c$ the entropy overtakes the energy $\rho$
 and above $T_c$  the free energies become exponentially degenerate. This means 
that the Z(N) symmetry is spontaneously broken.\footnote{This behaviour has been found analytically in gauge Potts models~\cite{potts} and in gauge theory by Monte Carlo simulations~\cite{kaj}. For other mechanisms of symmetry breaking at high temperature see ref.\cite{dvali} } 
 
The question is now: does the parameter $\alpha$ that controls the decay of the 
flux free energy correspond to a surface tension between two regions degenerate
 in energy? If so, there must be an order parameter telling the difference 
between the two degenerate states. 
 
\subsection{Domainwalls as Wilson line profiles}

To study domain walls one has to introduce the duals $Z_{\vec k}$ of the flux free 
energies. They are called ``twisted'' transition elements and are defined
by

\begin{equation}
Z_{\vec k}=Tr_{phys}\exp{(-H/T)}U_{\vec k}
\label{eq:twist}
\end{equation}

The $Z_{\vec k}$ are related to the $F_{\vec e}$ by the formula~\ref{eq:proj} for the 
electric projectors, that is, by a Z(N) Fourier transform. 

They can be rewritten into 4D ``twisted'' path integrals. To see how this works, take $U_{\vec k}=1$. This corresponds to the well-known pathintegral with an integration over periodic $A_0$ coming from the physicality constraints on the states. \footnote{Strictly speaking, $A_0$ need not be periodic in the Euclidean time
direction. Constraining  it to be periodic does not affect the thermodynamical properties} 

The presence of $U_{\vec k}$ {\it{does}} change the state the path integral starts with
at time $\tau=0$: it creates a  center group discontinuity when going  from 
one side ($z=0$) to the other ($z=L_z$). Then we go in the time direction
 to the point $\tau=1/T, z=L_z$, keeping this discontinuity. 
Going first in the time- and then in the z direction one meets no discontinuity, so 
we have a vortex in the  $z-\tau$ planes with strength $\exp{ik{2\pi\over N}}$. 
Consider a path ordered Wilson line
\begin{equation}
P\left(A_0(\vec x)\right)=P\{\exp{i\int_0^{{1\over T}}A_0(\tau,\vec x)d\tau}\}
\label{eq:line}
\end{equation}
 at $z=0$.
When we push the line to $z=L_z$ it will pick up the phase of the vortex,
as soon as it crosses the center of the vortex.\footnote{Clearly such a line in the z-direction will have this property too. However, it will not have a VEV at high T, contrary to the Wilson line. For a more complete discussion, see ref.~\cite{teper}}

This is  an important property of the line: at high T it will have a non-zero expectation value and change its phase going from one side of the box to the other. This answers the question we posed at the end of the last paragraph: the parameter $\alpha$ corresponds to a surface tension, the surface
separating two phases where the Wilson line takes different phases.
Note that we have not as yet a three dimensional interpretation of the Wilson line. As it stands it is a Euclidean path integral object.

The twisted transition element has been analysed numerically in its four
dimensional path integral form.~\cite{kaj} In three dimensions also the profiles $p(z)$ of the
Wilsonlines have been measured, clearly indicating domain walls above
$T_c$.~\cite{teper}

\subsection{Wilsonline and heavy quarksource}

We would like to associate the Wilsonline in the periodic system
with the presence of a heavy quark. This in order to  have a three
dimensional interpretation of the line and of the twisted transition element.

However in a periodic volume with Gauss' law everywhere imposed a single heavy quark source cannot exist.
So we have to drop the periodicity and search for convenient boundary conditions, that replace our twisted box and introduce a localised wall.
In this article we will not explain this in detail, but the strategy
can be read off from a  simple Z(2) lattice gauge model.~\cite{bronoff} The model is the limit of an SU(2) gauge model
where all particles are made very massive by adjoint Higgs multiplets. This leaves only Z(2) centergroup transformations, as they still commute
with the adjoint Higgses.
 There  one can show that  at high temperature a domain wall is created with a profile of positive energy density. The rigour is the same as used in proving that the Ising
model can produce walls.

\section{Effective action, perturbative expansion, and infrared cut off}\label{sec:infra}

The surface tension $\alpha$ was computed in perturbation theory.~\cite{bhatta} Also the profile of the Wilson line. The computation leads in a natural way to the realisation that the loop serves as a gauge invariant infra red cut-off~\cite{debbio} in the effective action. This is the subject of this section.

The effective action $U$ for the Wilson line is defined for any given profile $p(z)$ as follows:

\begin{equation}
\exp{-{U(p)\over T}L_{tr}^2}=\int DA_{\mu}\delta(p-\bar P(A_0))\exp{-{1\over {g^2}}S(A)}
\label{eq:eff}
\end{equation}

To avoid clutter in eq.~\ref{eq:eff} we left out the z dependence. $\bar P(A_0)$ stands for the normalised average over the transverse directions of the Wilson line. Hence it need not be unitary. 

The effective action gives the twisted transition element~(\ref{eq:twist}) by integrating
over all  profile configurations with boundary conditions appropriate
to the twist. In the time direction the statistics of the boson fields imposes
periodicity, for fermions anti periodicity.

Let us note that the Wilson line~(\ref{eq:line}) transforms under periodic gauge transformations as an adjoint. So only the phases are gauge invariant.
Hence the loop $P$ in eq.~\ref{eq:eff} stands for $TrP$, and to get all
eigenvalues one should admit winding the loop several times (from 1 to N-1) before taking the trace.

$U(p)$ has a symmetry under gaugetransformations that are periodic in the time direction modulo a centergroup
element. Such a transformation indeed leaves the action and the measure invariant, but changes the Wilson line by that same phase. Hence we have:
\begin{equation}
U(p \exp{ik{2\pi\over N}})=U(p)
\label{eq:znsymm}
\end{equation}
The same stays true when we admit fields with no Z(N) charge.
But fields
that do carry Z(N) charge will see their boundary conditions in the time direction changed by the phase. Hence the relation~(\ref{eq:znsymm}) is
no longer true; it only holds approximately, the better the larger the masses
of those fields are.

In what follows we  consider the perturbative expansion of $U$.
This implies we work at very high T, where the VEV of the Wilsonline has modulus one,
so only phases:
\begin{equation}
p(z)=\exp{i2\pi C(z)}
\label{eq:p}
\end{equation}

The matrix $C$ is in the Cartan subspace of the Lie algebra of SU(N) and
contains all the eigenvalues $C_i,i=1....N$, with the constraint that they add up to zero.
So our perturbative expansion will be around $C$ as background.

U(C), the effective action, will take the general form of a kinetic and a potential part:

\begin{equation}
U(C)= \int dz\left( K(C)Tr(2\pi T{\partial\over {\partial z}}C)^2 +\pi^2 T^4 V(C)\right)
\label{eq:general}
\end{equation}
Both  K and V are Z(N) periodic just as $U(C)$ in eq.~\ref{eq:znsymm}, and will loose that property when there are Z(N) charged fields.

At large values of $L_{tr}$ the pathintegral over all profiles is dominated  by the extremum of $U(p)$. At large values of $\vert z\vert$ the wall will be determined by the behaviour of kinetic and potential terms at C=0. The behaviour
of  the wall in that region
is determined by the large distance behaviour of the correlation of the
 phase of the loop. Hence it will be dominated by the Debije mass, not by
the magnetic glueball mass, since the correlation is odd under CT~\cite{arnold}.

\subsection{Background field expansion}\label{subsec:back}

The expansion introduces fluctuation fields $Q_{\mu}$ around
the background $C$:
\begin{equation}
A_{\mu}=2\pi T C \delta_{\mu ,0}+g Q_{\mu}
\label{eq:Q}
\end{equation}

Now we expand the Wilson line around $C$. This gives:
\bea
P(A_0)&=&\exp{i2\pi C}+\int^{1\over T}_0 d\tau \exp{(i2\pi TC\tau)}igQ_0(\tau)\exp{\left(i2\pi T C(1/T-\tau)\right)}\nonumber \\
& &+\int_0^{1\over T} d\tau_1\int^{1\over T}_{\tau_1} d\tau_2\exp{(i2\pi TC\tau_1)}igQ_0(\tau_1)\exp{\left(i2\pi TC(\tau_2-\tau_1)\right)}\nonumber \\
& & \hspace{2.5cm} igQ_0(\tau_2)\exp{\left(i2\pi TC(1/T-\tau_2)\right)}+O(g^3)
\label{eq:flucP}
\eea
 
This expansion satisfies the constraint in the effective action up to
terms of order $g$. The delta function is then expanded around these terms.
Since $C$ is diagonal the trace of the order $g$ term tells us that static diagonal $\bar{Q_0}$ should not be integrated over. On the other hand expanding the action gives a linear term containing {\it{only}} static diagonal $\bar{Q_0}$. So $C$ is the correct minimum for the expansion.

We have to choose gauge fixing and will take it of the form~\cite{gocksch}:
\begin{equation}
S_{gf}={1\over {\xi}}Tr({1\over {\xi^{\prime}}}D_0Q_0+\partial_iQ_i)^2
\label{eq:gf}
\end{equation}
which reduces to usual background field gauge fixing for $\xi^{\prime}=1$.
$\xi^{\prime}\to 0$ gives us static background gauge, in which $Q_0$ decouples from the loop in eq.~\ref{eq:flucP}.  

So in general thermal fluctuations of the loop expectation value will be $O(g^2)$, and in static background gauge absent.~\cite{gocksch} This is important for the tunneling results of next section. The fluctuations will contribute to the effective action, and render it gauge independent.~\cite{belyaev}

Another remark concerns the propagators of the quantum fields $Q$: with our gauge choice the background field will enter the quadratic part of the action through the covariant time derivative: $TrQ_{\mu}.(-D_0^2-\vec{\partial}^2).Q_{\mu}$. The background  enters only through commutators. Let us denote the field with
all components zero, except the one on the row i and column j by $Q^{ij}$.
Then the inverse propagator will look for {\it{all}} polarisations like:
\begin{equation}
\left( (2\pi T)^2(n+C_i-C_j)^2+{\vec p }\right)^2
\label{eq:prop}
\end{equation}

This shows that the static configurations ($n=0$) are screened by the
phases, except where the differences are integer valued. This happens
precisely in centergroup values $\exp{i2\pi C}=\exp{ik{2\pi\over N}}$.
Fields diagonal in colour are not screened. From those diagonal fields the constraint in eq.~\ref{eq:eff} eliminates part. In other words the $U(1)^{N-1}$
subgroup (including for the SM QED) keeps a infra red problem.

The same is true for any particle species that has no Z(N) charge, e.g.
is in the adjoint representation. Actually for those species all centergroup elements look the same. This confirms on the perturbative level what we
found quite generally in the previous subsection.

For particles in the fundamental N-dimensional  representation the inverse propagators 
look like:
\begin{equation}
\left((2\pi T)^2(n+C_i)^2+{\vec p}\right)^2
\label{eq:propfun}
\end{equation}

for the i-th component. Now there is still screening in the centergroup elements.

\subsection{Walls and profiles in perturbation theory}\label{subsec:walls}

The effective action is the key quantity to compute. In the limit of 
very large $L_{tr}$  its extremum is going to dominate the path integral 
over profiles. For  the one dimensional
action in eq.~\ref{eq:eff} this extremum  is simply given by the equations of motion for the profile:
\begin{equation}
\sqrt{K(C)}2\pi T{\partial\over {\partial z}}C-\pi T^2\sqrt{V(C)}=0
\label{eq:propfun}
\end{equation}
The eigenvalues  $C_1=-C_2$ in SU(2) are parametrised by  $2C=C_1$, V(C) is periodic mod 1 in C, as is obvious from the discussion on Z(N) symmetry in the previous sections.
It is normalised to zero for C=0.
Solving this equation with the boundary conditions C=0 and C=1 (p=1 and p=-1) at $z=\pm\infty$ gives a profile, that is controlled at large $\vert z\vert$ according to the general arguments just above subsection~\ref{subsec:back} by the Debije
mass 
\begin{equation}
m_D={1\over 2}\lim_{C\to 0}\sqrt{V(C)\over{K(C)}}
\label{eq:debije}
\end{equation}
Let us see how this works in perturbation theory. 

 One can readily do
this in the background expansion and one finds to  two loop order~\cite{bhatta}:  
\bea
 K(C)=1/g^2(1+ g^2 K_2(C))\nonumber \\
V(C) =C^2(1-\vert C\vert)^2 -5{g^2\over{8\pi^2}}C^2(1-\vert C\vert)^2
\label{eq:effpert}
\eea

The kinetic term $K_2(C)$ behaves well for all C, except at small values, where it develops
a pole.  This is in contrast with our general expectation of the large z 
behaviour. It should be governed by the Debije mass in eq.~\ref{eq:debije}. This 
equation, together with eq.~\ref{eq:effpert}, tells us that the correction to
the lowest order result $\sqrt{N/3} gT$ for $m_D$ is $(1+O({g^2\over C}))$!
However the work of Rebhan~\cite{rebhan} shows that this correction is {\it{not}}
$O(g^2)$. It is {\it{larger}}, of order $g \log (1/g)$ !

\subsection{Postmortem}\label{subsec:mort}

Our seemingly disastrous result, the pole in the kinetic term, just indicates, that for small $C$, naive perturbation theory breaks down.
For small values of C, say  $O(g^2)$, the screening of the propagators in eq.~\ref{eq:prop} becomes
ineffective. For those values of the background one has to do first the integration over all non-static configurations, and obtain an effective 3D action $S_3$.~cite{ginsparg} 
This integration will induce terms of order $g^2$ (the one containing the lowest order Debye
mass term for the static fourth component of the quantum field) and higher.  

The new action will only depend on the static quantum fields,  the background C
and inherits the gauge fixing in eq.~\ref{eq:gf}.  It will receive contributions from the non-static part of the constraint. The latter is gauge dependent and absent in static background gauge. Like in the 4D calculation
they are the guarantee for gauge choice independence. 

We have calculated the Debije mass from this effective action.~\cite{debbio} In Feynman background gauge there is only one 3D diagram contributing. The result coincides with that of Rebhan.  

\section{Effective potential for the Standard Model and beyond}\label{sec:sm}

In this section the one and two loop potentials for the Standard model and
beyond are analysed. 

There are two interesting types of minima: the absolute minimum and the 
metastable minima.~\cite{dixit} It will turn out that the latter are very long lived 
on cosmological scales.

Let us first analyse the contribution to the potential of a generic particle species coupled through the covariant time derivative to its Wilson line phase $x_s$:
\begin{equation}
D_0(x)=2\pi T(n+x)
\label{species}
\end{equation}

The one loop result for a complex boson neglecting its mass is then:

\bea
V_b(x)&=&Tr_T D^2(x)-Tr_{T=0}D^2(x)\nonumber \\
      &=&{-\pi^2\over{45}}T^4+{2\over 3}\pi^2 T^4 x^2(1-\vert x\vert)^2  
\label{boson}
\eea

The trace at temperature T stands for the sum over Matsubara 
frequencies and a regulated integral over $3-2\epsilon$ dimensions.
We subtracted the T=0 contribution, and get, because of eq.~\ref{species}
a result which is periodic in x mod 1.
The contribution of a Dirac helicity doublet is simply:

\begin{equation}
V_f(x)=-V_b(x+{1\over 2})
\label{fermion}
\end{equation}

The minus sign comes from the fermion determinant and the shift from the 
anti-periodicity for fermions used in eq.~\ref{species}.

Note that the sum of the two is antisymmetric when x is shifted over
${1\over 2}$. So SUSY theories, though broken at high temperature,
have this discrete symmetry for every species.  

Now we have  to specify what x is in terms of the various Cartan group charges of our gauge group.

For the SM and SU(5) we are working in a four dimensional space of
phases. So our matrix C (eq.~\ref{eq:p}) can be conveniently described by an SU(5) matrix in which
the phases for colour SU(3) are given by q and r, of weak $SU_L(2)$ 
by t, and of weak hypercharge U(1) by s:

\begin{equation}
diag\big({q\over 3}+{r\over 2}-{s\over 3},{q\over 3}-{r\over 2}-{s\over 3},-{2q\over 3}-{s\over 3},{t\over 2}+{s\over 2},-{t\over 2}+{s\over 2}\big)
\end{equation}

So e.g. for the righthanded electron x equals $-{1\over 2}s$.

All what is left to do  is to sum over all particle species. We then normalise to zero by subtracting out the Planck free energies in eq.~\ref{boson} and eq.~\ref{fermion}.
The result is plotted for SU(3) in fig.1, and for 
the SM and SU(5) against the weak hypercharge phase in fig.2. 
The Higgs content of SU(5) was taken to be the 5 and the 24. For SO(10) the 16 and and 45.
For SUSY SO(10) we took the complexified 45 {\it{and}} 54 to guarantee a
good SU(5) limit.

\begin{figure}[ht]
\centering
\epsfig{figure=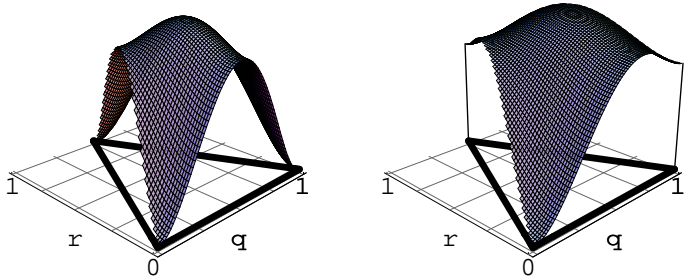,width=10cm}
\caption{$SU(3)$ potentials on the elementary cell (without and with six massless fermions).}
\label{fig:cell}
\end{figure}

For  SO(10) the plot in fig.2 is in terms of a U(1) charge u, orthogonal to  the SU(5) charges. The reason is that SO(10) contains SU(5)xU(1).

All minima are in the centergroup elements. The trivial centergroup element contains the absolute minimum. The other centergroup elements are
metastable points ( though not all, as in the SM). There would have been degeneracy, had all the species been Z(N) neutral. But the fermions,
the fundamental Higgs have Z(N) charge, and lift the degeneracy.

How can we be sure of seeing all the relevant minima? This will be 
explained in subsection~\ref{subsec:where}.

Then there are the tunneling rates of these metastable states into
the stable state. They will be discussed in  subsection~\ref{subsec:tunnel}.
Lastly the thermodynamics is discussed.  

\begin{figure}[ht]
\centering
\epsfig{figure=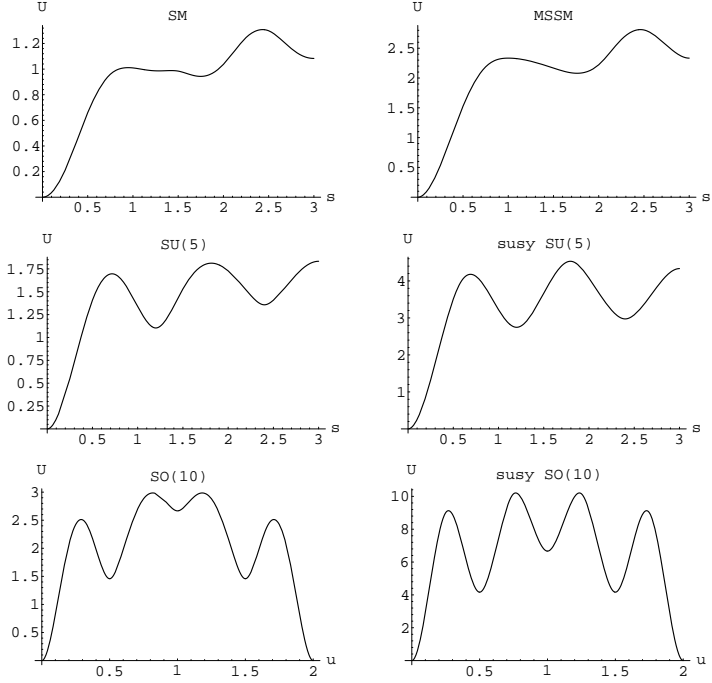,width=10cm}
\caption{Potentials in the weak hypercharge direction $s$ ($u$ for $SO(10)$).}
\label{fig:hyper}
\end{figure}

\subsection{Centergroup lattice of SU(N) and SO(2N) groups}\label{subsec:where}

We begin with the lattice of points in the Cartan subspace, that contains
all centergroup elements. Such a lattice is easy to generalise from the
the case of SU(3) in fig.1. 

Consider the set of N basis vectors $e_1,e_2, ....,e_N$ given by the 
diagonal NxN matrices $e_1={1\over N}diag(1,1,.....,1,1-N),e_2=
{1\over N}diag(1-N,1,....1),....,e_N={1\over N}diag(1,1,....,1-N,1)$. Their sum adds up to zero.
Taking any linear combination L of the $e_k$ with integer coefficients
will give a centergroup element $\exp{i2\pi L}$. The inverse is true too:
all centergroup elements are to be found on this lattice.

This lattice contains many elementary cells on which the potential is identical. This is because of the symmetries in eq.~\ref{species}.

A convenient cell is starting in 0, and formed by the succession $e_1,
e_1+e_2,.....,e_1+e_2+......+e_{N-1}$. Inside this cell there are
conjugated points $e_1$ and $e_1+e_2+....+e_{N-1}$, $e_1+e_2$ and
$e_1+e_2+....+e_{N-2}$, etc. related by charge (or CP) conjugation,
and with complex conjugate values for the centergroup elements.

The cell of SO(10) is related to that of SU(5) because SO(10) or rather
its covering group Spin(10) contains SU(5)xU(1). 
The centergroup is Z(4) as one can easily check from the 16 dimensional
spin representation. We normalise the u variable by fixing P=-1 at u=1
and all other phases zero. 
Consider the cell of SU(5). Shift from $e_1$ by  $-{1\over 5}$ in the u 
direction to get the vector $u_1$ with P=-1. Shift $e_1+e_2$ 
by ${1\over {10}}$ in the u direction to get $u_2$ with P=i. The complex conjugate of $u_2$ is $u_3$ and is obtained from $e_1+e_2+e_3$ by shifting
over $-{1\over {10}}$, and that of $u_1$ from $-e_5$ by shifting
over ${1\over 5}$. This generalizes easily to $SO(2N)$.

Where are the minima of the potential? If only Z(N) neutral fields
are around there is a proof to all orders in perturbation theory
that the minima are in the centergroup elements~\cite{gocksch}.

As discussed above, fermions and Higgs fields lift the degeneracy.
We have verified numerically that no other metastable points develop.
So we have the following working hypothesis:

{\it{Any metastable or stable minimum must be in the centergroup.}}

A last question concerns how the minima of say SU(5) do give rise to
the minima in the standard model, when decoupling the heavies in SU(5).

Our working hypothesis will prove useful here. The heavies in SU(5) are given a common mass m, that we switch on from its value 0 in SU(5) to
$\infty$ in the SM. They are all multiplets of the SM gauge group,
so switching on m will cause a movement of the minima of SU(5) along
the centergroup elements of the SM, to wit Z(3)xZ(2)xU(1). Since the 
process is continuous the minima can only slide along the six U(1)
lines corresponding to the discrete Z(3)xZ(2) group. This is shown in fig.3 for the $U(1)$ line $q=r=t=0$. 

\begin{figure}[hb]
\centering
\epsfig{figure=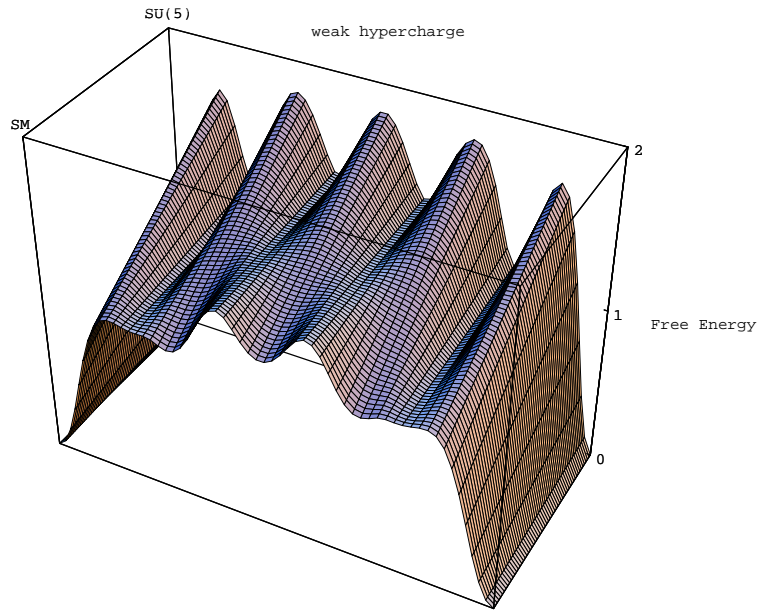,width=8.5cm}
\caption{Potential interpolating between $SU(5)$ and SM.}
\label{fig:decoupl}
\end{figure}

Similarly the evolution of the SO(10) minima into SU(5) minima
occurs along the 5 U(1) lines, that we used in the construction of the
SO(10) cell out of the SU(5) cell.

\subsection{Tunneling paths and tunneling rates}\label{subsec:tunnel}

Semi classical methods are appropriate, in our situation with 
small coupling, to compute the tunneling rate $\Gamma$ per unit volume:  
\begin{equation}
\Gamma=aT^4\exp{-S_b}
\label{life}
\end{equation}

The dimensionless quantity $a$ derives from the determinant of
the fluctuations around the classical bounce configuration B (for QCD and
the SM this was done in ref.~\cite{ignat})

The computation of the bounce consists of finding the solutions to
the Euclidean equations of motion. The temperature T is much larger than the scale of the 4D Euclidean bounce, we can do the calculation in 3D~\cite{linde} for a radially symmetric bounce:
\begin{equation}
{d^2B\over{{dr}^2}}+{2\over r}{dB\over{dr}}=-V^{\prime}(B)
\label{eq:bounce}
\end{equation}
Boundary conditions are $\lim_{r\to\infty}=B_{ms}$ and ${dB\over{dr}}=0$
at the starting point $r_0$. Our bounce is a multicomponent object
(4 or 5) and $B_{ms}$ is the metastable point in the elementary
cell. We study the decay to the stable vacuum at the origin, so 
$r_0$ is near zero, depending on the thickness of the wall. We have  developed a numerical method that solves the equations of motion and can
handle the multicomponent field. 

Luckily there is a lot of symmetry in our problem, that allows an educated guess of the actual tunneling path. So we can reduce the problem to
a one component bounce, solvable with the undershoot-overshoot method.

The bounce path from a given  metastable point in the cell is simply given
by the straight path from that point. The direction of the bounce path is always  inside a valley of the potential except for the SM where it stay near to a valley. This was corroborated by our numerical method for the simple groups. For the SM the deviation from a straight path was small but significant \footnote{
Not all metastable minima in fig. 2 lie in the same cell as one can easily
check.}. 
The results for the bounce actions are presented in table 1.

\begin{table}[ht]
\caption{Numerical data on the bounces.\label{tab:exp}}
\vspace{0.2cm}
\begin{center}
\begin{tabular}{|l||c|c|c|}
\hline
& metastable points & Bounce action           & Critical radius  \\
&                   &$(S_b/{4\pi^2\over g^3})$&$(R/gT)$\\
\hline \hline
SM           & $s=1.28, q=r=t=0$          & 6.99       & 2.7   \\
\cline{2-4} 
             & $q=1 ,s=-0.25, r=t=0$       & 2.64      & 2.0 \\
\hline
   MSSM         & $s=1.761, q=r=t=0$      & 37.60     & 3.25 \\
\hline
SU(5)        & $p=\exp{(i4\pi/5)}$ & 10.63      &  2.3   \\
\cline{2-4}
             & $p=\exp{(i2\pi/5)}$ & 86.63      &  5.3   \\
\hline
susy SU(5)   & $p=\exp{(i4\pi/5)}$  &  9.92          &  1.75   \\
\cline{2-4}
             & $p=\exp{(i2\pi/5)}$ &    50.47   &  3.3  \\
\hline
SO(10)       &$p=-1$               &  20.99    &  2.75   \\
\cline{2-4}
             &$p=i$                &  308.8     &  7.4   \\
\hline
susy SO(10)  & $p=-1$                    &   36.87         &  2.2   \\
\cline{2-4}
             & $p=i$                    &  286.07          &  5.1   \\
\hline

\end{tabular}
\end{center}
\end{table}

What is striking is the large value of all bounce
actions. Tunneling temperatures $T_{tun}$ can be estimated by the following simple-
minded argument. Tunneling occurs at a time $t_c$ determined by $\Gamma t^4_c\sim 1$. Matter in our metastable states has qualitatively the same free energy
as in the stable vacuum. Only the pressure $-\pi^2T^4 V(C_{ms})$ is lower
in the metastable states. So we will assume the relation $t\sim M_{Planck}/T^2$. Then $T_{tun}\sim M_{Planck}\exp{-S_b/4}$ , using eq.~\ref{life} to eliminate $t_c$. Putting in the numbers from the table shows that the metastable states easily survive all known cosmological transitions. 

It means that the behaviour of the potential as function of the Higgs fields 
will be very relevant for how the metastable states will decay.  This is the subject of section~\ref{sec:higgs}.

Fig. 4 shows the effect of the two loop contribution. the couplings were taken to be $\alpha=\beta=\lambda_1=\lambda_2=0.1$ and $g=0.4$.
The sign of the contribution is uniformly negative. That this was the case at
$C=0$ was known since long~\cite{kapusta}.

\begin{figure}[ht]
\centering
\epsfig{figure=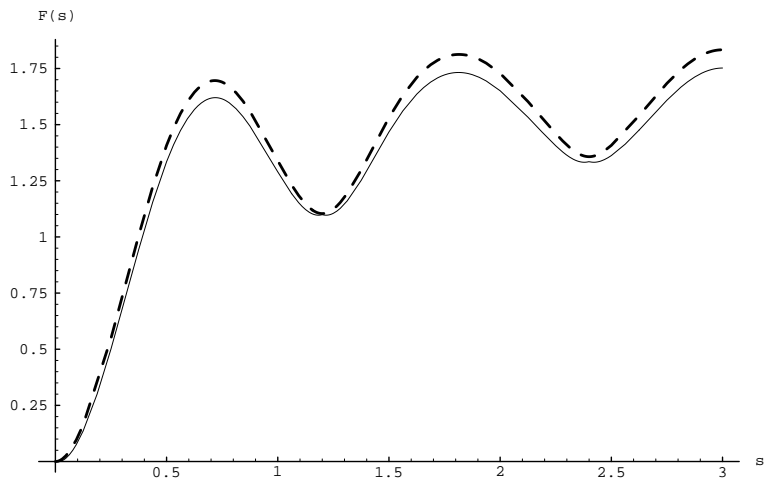,width=10cm}
\caption{$SU(5)$ potential: one-loop (dashed) and two-loop (solid).}
\label{fig:twoloop}
\end{figure}

\subsection{Thermodynamics}

The calculation of the surface tension~\cite{bhatta} motivated a renewed interest
in the thermodynamics of the minima. In all the models considered in the previous subsections the energy and entropy densities of the minima are positive. 
But the thermal boundary conditions on fields with non-trivial Z(N) charge
are changed in the metastable minima by a Z(N) phase, and as a consequence the occupation number is changed~\cite{semenoff}:
\begin{equation}
n(E)={1\over{\exp{({E\over T}+ik{2\pi\over N})}\pm 1}}
\label{density}
\end{equation}

So in the case of fermions one finds a purely imaginary value for the
fermion number. So far no reasonable interpretation has been found for this behaviour. States with complex conjugate Z(N) phase are related by CP
conjugation. Our phases have the C and P transformation properties of a diagonal
$A_0$, so are CP odd. This singles out the self conjugate states like the
one in SO(10) with $P=-1$. There, fermion number is zero.

Let us finally mention a case without any thermodynamic anomalies: large N pure gauge theory. To see this, consider the free energy without the Planck  contribution:
\begin{equation}
V(C)={4\over 3}\pi^2 T^4\sum_{1\le i\le j\le N} C_{ij}^2\left(1-\vert C_{ij}\vert\right)^2
 \label{eq:largen}
\end{equation}

\noindent where $C_{ij}=C_i-C_j$. Due to the permutation symmetry of the variables $C_i$, this potential has an extremum in the barycenter of the elementary cell. in the one loop approximation this is an absolute maximum.
Let us compute its value $V_{max}$. In the barycenter the value of $C_{ij}$ is
${i-j\over N}$ . Substituting in eq.~\ref{eq:largen} gives:
\begin{equation}
V_{max}={\pi^2 T^4\over {45}}(N^2-1)\left(1+O({1\over {N^2}})\right)
\label{maxlargen}
\end{equation}

To leading order we find precisely the Planck free energy, but with the opposite sign! Hence adding the Planck free energy  we have positive
energy and entropy density, except in the maximum where both are zero.
This result remains true in the next order, since it gives positive contribution.~\cite{kapusta} 
 
For finite N there is a small cap around the maximum, where both are negative.
 
\section{Combining Higgs and Wilson line potential}\label{sec:higgs}
 
In this section we will compute the classical and part of the one loop contribut
ion to the mass term in the Higgs-Wilson line potential. Already on the
classical level
a few subtleties merit attention.

\subsection{Classical effective action}\label{subsec:classhiggs}
 
 It is best to illustrate the problem by the specific examples of the adjoint
Higgs field $\Sigma$ and the fundamental Higgs field $H$.
 
The terms of interest in the original 4D action are the kinetic terms of the 
two Higgses, minimally coupled to the gauge fields:
\begin{equation}
S=\int d\vec x d\tau [Tr(D_{\mu}\Sigma)^2 + (D_{\mu}H)^{\dagger}D_{\mu}H+....]
\label{higgskin}
\end{equation}
 
To get the 3D effective action one would guess that the classical contribution  would just be the 3D reduction of the kinetic terms. That would give us terms like $Tr[C,\Sigma]^2$.
 
This can{\it{not}} be correct. We know already on general grounds that the effective action for Z(N) neutral fields like $\Sigma$ is identical in all centergroup elements. But this commutator term is not.
 
What couples in the 3D action to the Higgs is the phase of the loop p(C).
We define the matrix $\log p(C)$ through the diagonal form of the argument, and under a gauge transform $\Omega$ we have
\begin{equation}
\log p(C^{\Omega})=\Omega \log p(C) \Omega^{\dagger}
\label{transpC}
\end{equation}
 
Then the coupling of the Higgses to the Wilson line is given in a gauge invariant form in the effective action $U(C)$ by:
\begin{equation}
U(C)=\int dz T^2[Tr[Im \log p(C),\Sigma]^2 + (Im \log p(C)H)^{\dagger}(Im \log p(C)H)+....
\label{higgsclasspot}
\end{equation}
 
The VEV of the Wilson line induces mass terms that merit comment:
 
i)The adjoint mass term disappears in the centergroup, since the commutator vanishes there.
 
ii)The mass term for the fundamental Higgs does not vanish in the centergroup.
Define $\vert \Im\log p(C)\vert\le\pi$, then the mass term is the same for conjugate vacua.
 
iii)Outside the centergroup the induced Higgs masses are not necessarily the
same, i.e. SU(5) invariant. In general, the potential is not anymore function of only $Tr\Sigma^2$
etc., but contains other invariants involving $p(C)$.
 
\subsection{Quantumcorrections to the Higgs potential}\label{subsec:quantumhiggs}
 
The quantum corrections are crucial to the understanding of the temperature dependence of the Higgs potential and hence for the occurrence of a phase transition. These corrections have been calculated for the trivial Wilson line condensate.
 
In the presence of a non-trivial condensate these corrections will be described 
below.
 
Let us take the concrete example of SU(5). The Higgs potential in the condensate characterised by $\exp{ik{2\pi\over 5}}$ takes the form:
\bea
\pi^2 T^4V&=&M^2(T,k)Tr\Sigma^2 +m^2(T,k)H^{\dagger}H +\lambda_1(Tr\Sigma^2)^
2+\lambda_2Tr\Sigma^4\nonumber \\
             & &+\alpha H^{\dagger}HTr\Sigma^2+\beta H^{\dagger}\Sigma^2H
\label{higgswilsonpot}
\eea
 
The mass terms are temperature and k dependent through loop corrections. In the 
one loop calculation that we completed the $\Sigma$ mass has the following
corrections in Landau gauge:
 
\bea
M^2(T,0)=\left( 104 \lambda_1 + {188\over 5} \lambda_2 +60 g^2 + {20} \alpha + {4} \beta \right) {T^2\over 24}
\eea

\bea
M^2(T,1)=\left( 104 \lambda_1 + {188\over 5} \lambda_2 +60 g^2 - {36\over 5} \alpha - {36\over 25} \beta \right) {T^2\over 24}
\eea
 
The mass term contains a dependence on the condensate uniquely through
the couplings to the fundamental Higgs. This is a consequence of the $\Sigma$
propagators not depending on the condensate. The induced terms will not 
change much the Higgs potential, so the SU(5) transition in a non-trivial condensate will not change essential features.
 
On the other hand the transition in SO(10) and the electro weak are changed
by  induced mass terms of order one:
\begin{equation}
m^2(T)=-m^2+ T^2(Im \log p(C))^2+ O(couplings)
\label{eq:funhiggsmass}
\end{equation}
 
\subsection{Some physical consequences}\label{subsec:phys}

Striking features of the potential are:

i) CP conjugated states in the SM.

ii) Occurrence of thermal mass terms of order unity for Higgs fields with
non-trivial Z(N) charge.

The first point has been discussed~\cite{watson} and we will not go into it. they involve {\it{pe se}} states with imaginary fermion number.
 
The second point has two important consequences. First, the transition will be at a {\it{lower}} temperature~cite{lee} $T_c(ms)$ than in the stable state at $T_c$ by a factor $\sqrt {\hbox{couplings}}$.
This follows by equating the temperature dependent mass tem in eq.~\ref{eq:funhiggsmass} to zero at the critical temperature.
 
Second, before the transition takes place, the system will undergo thermal inflation.~\cite{infla} This happens in both states with complex and real values for the Wilson line in SO(10).
 
Let us denote the VEV of the SO(10) Higgs H in the 16 representation
by $v$. The relevant terms in the effective action are then:
\begin{equation}
\pi^2T^4V={1\over 2}{m^2\over {v^2}}[H^{\dagger}H-v^2)^2 +T^2(k\pi)^2H^{\dagger}
H+...
\label{eq:so10}
\end{equation}
 
Here $k=1$, or $\pm{1\over 2}$ according to wether $p=-1$, or $\pm i$.
As long as the temperature is above the value $T_{out}={m\over{k\pi}}$ we have no
VEV for the Higgs, and the energy density $\epsilon$ of the vacuum equals ${1\over 2
}(mv)^2$. This defines a second temperature scale $T_{in}=(mv)^{1\over 2}$. 
Below this scale the thermal contribution will become small with respect to
the vacuum  energy density till we reach the lower scale $T_{out}$. Below that scale
the symmetry breaks, the Higgs gets the VEV $v$ and and the energy density becomes proportional to $T^2$, from eq.~\ref{eq:so10}. So in between 
these
temperatures we have inflation. 
 
There is a third scale, appearing in the Friedman equation coupling the radius R
 of a flat universe to its energy density:

\begin{equation}
{1\over R^2}\left({dR\over dt}\right)^2={\epsilon(T)\over {M_{Pl}}^2}\nonumber\\
\label{fried}
\end{equation}
 
During the inflationary period the right hand side of this equation 
is approximately constant and equals  the square of ${v\over M_{Pl}m}$.
This scale is  the Hawking-Gibbons  temperature $T_{HG}$ and lies {\it{below}} $
T_{out}$.
That is to say, it will not play any role in the inflationary period.
This is a good feature, because $T_{HG}$ is a lower bound on the temperature during inflation.
  
\section{Conclusions}
 
We have given an overview of the use of the Wilson line in high 
temperature gauge field physics.
 
The effective potential of the Wilson line does not only give rise to 
symmetry
breaking at high temperature. It provides in a natural way a gauge 
invariant
infra red cut off and is useful in establishing the perturbative series 
for
static quantities like free energy and  Debije screening length. Many applications are possible, e.g. Callan-Symanzik type equations for the 
infrared behaviour of hot gauge theory~\cite{buffa}. Very striking are its universal features: Long lived metastable states with potentialy very interesting consequences.
 
\section*{Acknowledgements}
Our thanks go to Graham Ross, Misha Shaposhnikov and Mike Teper for insisting on various questions and criticisms. S.B. acknowledges an Allocation de Recherche MENESR. C.P.K.A. thanks the organizers of this meeting for their invitation and wonderful hospitality.

\end{document}